\documentclass[sn-nature]{sn-jnl}

\usepackage{amsmath,amssymb,amsthm,amsfonts,mathrsfs,mathtools,bm}
\usepackage{lmodern,textcomp,manyfoot}
\usepackage{multirow,booktabs}
\usepackage{pgf,graphicx}

\newcommand{\ns}{\text{\tiny ns}}

\begin{document}
\title[Dynamic Factor Analysis of Price Movements in the Philippine Stock Exchange]{Dynamic Factor Analysis of Price Movements in the Philippine Stock Exchange}

\author*[1]{\fnm{Brian Godwin} \sur{Lim}}\email{lim.brian\_godwin\_sy.la6@is.naist.jp}

\author[1]{\fnm{Dominic} \sur{Dayta}}

\author[1]{\fnm{Benedict Ryan} \sur{Tiu}}

\author[1,2,3]{\fnm{Renzo Roel} \sur{Tan}}

\author[4,5]{\fnm{Len Patrick Dominic} \sur{Garces}}

\author[1]{\fnm{Kazushi} \sur{Ikeda}}

\affil[1]{\orgdiv{Division of Information Science, Graduate School of Science and Technology}, \orgname{Nara Institute of Science and Technology}, \orgaddress{\postcode{630-0101}, \state{Nara}, \country{Japan}}}

\affil[2]{\orgdiv{Department of Applied Mathematics and Physics, Graduate School of Informatics}, \orgname{Kyoto University}, \orgaddress{\postcode{606-8501}, \state{Kyoto}, \country{Japan}}}

\affil[3]{\orgdiv{Department of Quantitative Methods and Information Technology, John Gokongwei School of Management}, \orgname{Ateneo de Manila University}, \orgaddress{\postcode{1108}, \state{Metro Manila}, \country{Philippines}}}

\affil[4]{\orgdiv{School of Mathematical and Physical Sciences, Faculty of Science}, \orgname{University of Technology Sydney}, \orgaddress{\postcode{2007}, \state{New South Wales}, \country{Australia}}}

\affil[5]{\orgdiv{Department of Mathematics, School of Science and Engineering}, \orgname{Ateneo de Manila University}, \orgaddress{\postcode{1108}, \state{Metro Manila}, \country{Philippines}}}

\abstract{
The intricate dynamics of stock markets have led to extensive research on models that are able to effectively explain their inherent complexities. This study leverages the econometrics literature to explore the dynamic factor model as an interpretable model with sufficient predictive capabilities for capturing essential market phenomena. Although the model has been extensively applied for predictive purposes, this study focuses on analyzing the extracted loadings and common factors as an alternative framework for understanding stock price dynamics. The results reveal novel insights into traditional market theories when applied to the Philippine Stock Exchange using the Kalman method and maximum likelihood estimation, with subsequent validation against the capital asset pricing model. Notably, a one-factor model extracts a common factor representing systematic or market dynamics similar to the composite index, whereas a two-factor model extracts common factors representing market trends and volatility. Furthermore, an application of the model for nowcasting the growth rates of the Philippine gross domestic product highlights the potential of the extracted common factors as viable real-time market indicators, yielding over a 34\% decrease in the out-of-sample prediction error. Overall, the results underscore the value of dynamic factor analysis in gaining a deeper understanding of market price movement dynamics.
}


\maketitle

\section{Introduction}
The financial markets have been the subject of interest for researchers, practitioners, and investors as they offer opportunities to invest excess funds and generate positive returns. Among the primary objectives in the study of financial markets is the accurate prediction of stock price movements as this allows one to outperform the market and achieve significant gains. Toward this goal, substantial efforts have been dedicated to develop models that can effectively capture the complex dynamics of the stock markets, providing tools for making informed decisions based on historical data \cite{CHERNOV2003225,KOTHARI1995155,LONG2019163}. 

In recent years, machine learning models have gained popularity for the selection and extraction of features \cite{Kumari2023,htun_survey_2023} and for the prediction of future price movements \cite{Muhammad2023,li_clustering-enhanced_2023,lawi_implementation_2022,shen_short-term_2020,zhong_predicting_2019}. Despite their success, the \textit{black box} nature of these models hinders their interpretability. On the other hand, traditional asset pricing models, such as the capital asset pricing model, the Fama-French models, and the arbitrage pricing theory model, offer a simpler framework by characterizing the linear relationships between stock returns and some underlying factors \cite{Giglio2021,lintner_valuation_1965,sharpe_capital_1964,fama_common_1993}. However, these interpretable linear models often fall short in capturing the complexities of market phenomena. Thus, to provide a deeper understanding of stock price movement dynamics, models must combine the interpretability of conventional linear asset pricing models with the predictive capabilities of machine learning models.

To this end, the paper leverages econometrics literature by analyzing price movements in the stock market from the lens of dynamic factor analysis. Although not commonly used in financial applications, the dynamic factor model (DFM) may be used to explain stock returns as the sum of a common component and an idiosyncratic component \cite{DFM}. The former is further decomposed as a linear combination of a set of predictive features, called the common factors, extracted by the model in an unsupervised manner. The DFM thus offers an interpretable way to explain market phenomena, similar to traditional linear asset pricing models, while possessing the predictive capabilities of machine learning models. Moreover, while the model has been predominantly used for prediction, the paper focuses on analyzing the extracted loadings and common factors to provide an alternative perspective on understanding the complex dynamics of price movements in the stock market.

Using Kalman methods and maximum likelihood estimation, the analysis of the Philippine Stock Exchange, validated against the capital asset pricing model, provides new and alternative insights into classical market theories. Notably, the common factor in a one-factor model may be used to represent the systematic or market dynamics similar to the composite index while the common factors in a two-factor model may be used to represent market trend and volatility. Moreover, an application on nowcasting the Philippine gross domestic product growth rates further demonstrates the utility of the extracted common factors as viable real-time market indicators, achieving more than 34\% reduction in out-of-sample prediction error. These results highlight the unique perspective of dynamic factor analysis in understanding the dynamics of price movements in the market.

The paper is organized as follows. The second section describes related works with a particular focus on the capital asset pricing model, the arbitrage pricing theory model, and the principal component analysis. The succeeding section formally introduces the dynamic factor model which addresses the limitations of the previous models. The section also covers the model fitting methodology, the model validation procedures, and the model implementation details. The fourth section then presents the results of the model when applied to the Philippine Stock Exchange. This is followed by a demonstration of the utility of the extracted common factors in a macroeconomic nowcasting application. Finally, the last section concludes with a summary of the research and recommendations for future work.

\section{Related Works}
Recent developments in the analysis of price movements in stock markets have largely centered on the application of machine learning models, which generally serve two main objectives---performing feature extraction for downstream analysis tasks and predicting future price movements. Toward the first objective, models like random forests and autoencoders have been employed to extract features for explaining price movements \cite{Kumari2023,htun_survey_2023}. For example, \cite{gunduz2021efficient} applied variational autoencoders to extract features that successfully predicted the direction of stock price movements using long short-term memory and LightGBM models. Additionally, \cite{shahvaroughi2021forecasting} utilized genetic algorithms to select representative features for the same purpose with a simple neural network model. On the other hand, toward the second objective, \cite{wang2024stock} demonstrated the effectiveness of a neural network model in capturing the non-linear relationships that firm-specific and macroeconomic factors have on stock price returns. Similarly, \cite{htun2024forecasting} explored the use of random forests, support vector machines, and long short-term memory models to predict the excess return of a stock relative to a composite index. Other studies have followed a similar approach, employing deep learning models such as transformers and gated recurrent units to capture the complex dynamics of stock price movements \cite{Muhammad2023,li_clustering-enhanced_2023,lawi_implementation_2022,shen_short-term_2020,zhong_predicting_2019}. It is worth mentioning that most of these models rely only on historical stock price data as features. Nevertheless, some of these models try to incorporate additional features like technical indicators \cite{gunduz2021efficient,shahvaroughi2021forecasting}, fundamental indicators \cite{wang2024stock,shen_short-term_2020}, and macroeconomic indicators \cite{wang2024stock} to boost predictive capability. Despite their superior performance, these machine learning models often lack the interpretability of classical linear models. Moreover, the features extracted through these models may not be meaningful for analyses beyond predicting price movements.

The dynamic factor model from econometrics literature addresses these limitations of machine learning models while remaining performant in predictive applications \cite{luciani_nowcasting_2018,hayashi_nowcasting_2022,chernis2020three}. Notably, this linear factor model also provides interpretable loadings and latent features that may be used for further analysis. As a prerequisite for introducing the model, the capital asset pricing model and the arbitrage pricing theory model are first introduced, offering insights into the traditional methods for analyzing stock price movements. Following this, the principal component analysis is presented as an alternative method for extracting data-driven factors for the arbitrage pricing theory model.

\paragraph{Capital Asset Pricing Model}
The capital asset pricing model (CAPM) is widely recognized in finance literature to explain stock price returns. It describes the linear relationship between the expected return of a given stock and its exposure to systematic or market risks \cite{lintner_valuation_1965,sharpe_capital_1964}. Suppose $\mathbb{E}\left(R_i\right)$ is the expected return of stock $i$. The model assumes the following dynamics
\begin{equation}
    \mathbb{E}\left(R_i\right) - R_F = \beta_i \left[\mathbb{E}\left(R_M\right) - R_F\right], \label{eqn:capm}
\end{equation}
where $R_F$ is the risk-free rate of return, $\mathbb{E}\left(R_M\right)$ is the expected market return, and $\beta_i$ is the CAPM beta of stock $i$ that measures the sensitivity of the risk premium $\mathbb{E}\left(R_i\right) - R_F$ to the expected excess market return $\mathbb{E}\left(R_M\right) - R_F$.

Numerous studies use CAPM to investigate the relationship between risk and return \cite{blume_new_1973,perold_capital_2004,elbannan_capital_2014,rossi_capital_2016}. It assumes that the expected return of a stock co-moves with the expected return of the market, and that variations in the CAPM beta are sufficient to explain the cross-sectional differences in stock price returns.

\paragraph{Arbitrage Pricing Theory Model}
The arbitrage pricing theory (APT) model is another linear model widely used in finance literature to explain stock price returns. It extends the CAPM since empirical evidence suggests the need for a multi-factor model to explain stock price dynamics \cite{barucci_factor_2017}. The APT model assumes that stock returns are explained by a linear combination of a finite number of risk factors and a random factor specific to each stock. Suppose $R_i$ is the return of stock $i$. The model assumes the following dynamics
\begin{equation}
    R_i - R_F = \boldsymbol{\beta_i}^\top \boldsymbol{F} + Z_{i}, \label{mod:apt}
\end{equation} 
where $\boldsymbol{F}$ is a vector of $n$ risk factors, $Z_{i}$ is a stock-specific random factor for stock $i$, and $\boldsymbol{\beta_i}$ measures the sensitivity of stock $i$ to the risk factors. 

The APT model relaxes some assumptions of the CAPM and uses firm-specific or macroeconomic factors for $\boldsymbol{F}$ to explain the stock price returns. Firm-specific factors typically include book-to-market ratio, dividend yield, and cash-flow-to-price ratio while macroeconomic factors usually include expected inflation, yield spread between long-term and short-term interest rates, and yield spread between corporate high-grade and low-grade bonds \cite{barucci_factor_2017}. With the inclusion of different risk factors, the model can better explain stock price movements than CAPM \cite{reinganum_arbitrage_1981,elshqirat_empirical_2019}.

Despite these advantages, the question of which and how many factors to include remains unresolved, with empirical evidence suggesting that models utilizing derived factors may sometimes outperform those based on traditional economic and financial indicators \cite{french2017,reinganum1981arbitrage}. This makes a compelling case for using derived factors from models like principal component analysis, which provides a systematic and data-driven approach to factor extraction, addressing the challenge of factor selection within the APT model.

\paragraph{Principal Component Analysis}
The principal component analysis (PCA) is another linear model widely used in literature. Unlike the CAPM and APT model, which are used to explain stock price returns, PCA is primarily used for dimensionality reduction, compressing high-dimensional data into a lower-dimensional representation that preserves as much information and variability from the original data as possible \cite{PCABook,PCAOrig}. The compressed data, known as principal components, are mutually uncorrelated linear combinations of the original variables \cite{jolliffe_principal_2016}. Suppose the data is represented as a $T \times S$ matrix $\boldsymbol{X} = \left[\boldsymbol{x_1}, \boldsymbol{x_2}, \ldots, \boldsymbol{x_S}\right]$ containing $T$ observations of $S$ variables. The model determines the linear combination 
\begin{equation}
    \boldsymbol{Xa} = \sum_{s=1}^S a_s \boldsymbol{x_s} \label{eqn:pca}
\end{equation}
that maximizes the variance given by $\text{Var}\left(\boldsymbol{Xa}\right) = \boldsymbol{a^\top \hat{\Sigma} a}$, where $\boldsymbol{\hat{\Sigma}}$ is the sample covariance matrix of $\boldsymbol{X}$. The problem thus reduces to maximizing $\boldsymbol{a^\top \hat{\Sigma} a}$ subject to $\lVert\boldsymbol{a}\rVert = 1$, restricting $\boldsymbol{a}$ to be a unit vector. Using Lagrange multipliers, this is equivalent to maximizing
\begin{equation}
     \boldsymbol{a^\top \hat{\Sigma} a} - \lambda \left(\boldsymbol{a^\top a} - 1\right).
\end{equation}
The optimization problem above results in the equation $\boldsymbol{\hat{\Sigma}a} = \lambda \boldsymbol{a}$, indicating that $\boldsymbol{a}$ is a unit eigenvector and $\lambda$ is the corresponding eigenvalue of the sample covariance matrix $\boldsymbol{\hat{\Sigma}}$. Moreover, since
\begin{equation}
    \text{Var}\left(\boldsymbol{Xa}\right) = \boldsymbol{a^\top \hat{\Sigma} a} = \lambda \boldsymbol{a^\top a} = \lambda,
\end{equation}
$\lambda$ must be the largest eigenvalue of $\boldsymbol{\hat{\Sigma}}$. The first principal component is therefore calculated as $\boldsymbol{Xa_{(1)}}$, where $\boldsymbol{a_{(1)}}$ is the unit eigenvector associated with the largest eigenvalue of $\boldsymbol{\hat{\Sigma}}$. The succeeding principal components may be obtained similarly by adding the constraint 
\begin{equation}
    \text{Cov} \left(\boldsymbol{Xa}_{\left(i\right)}, \boldsymbol{Xa}_{\left(j\right)}\right) = \boldsymbol{a_{\left(i\right)}^\top \hat{\Sigma} a_{\left(j\right)}} = \lambda_{\left(j\right)} \boldsymbol{a_{\left(i\right)}^\top a_{\left(j\right)}} = 0
\end{equation}
or equivalently, $\boldsymbol{a_{(i)}^\top a_{(j)}} = 0$ for $j < i$. This results in $\boldsymbol{Xa_{(i)}}$ being the $i$th principal component, where $\boldsymbol{a_{(i)}}$ is the unit eigenvector associated with the $i$th largest eigenvalue of $\boldsymbol{\hat{\Sigma}}$ \cite{PCABook}. 

PCA is widely used in both literature and practical applications \cite{ghorbani2020stock,lim2024_dpca,yu2023engineering,xi2024factor}. It learns the optimal linear compression of high-dimensional data into principal components without requiring additional data, making it an unsupervised model for feature extraction. However, PCA inherently assumes that the $T$ observations are independent, which does not hold when $\boldsymbol{X}$ is a time-series data. 

To address the limitations of these models, the paper considers the dynamic factor model which integrates the strengths of both linear models and machine learning models, effectively balancing interpretability and predictive capability. In particular, while the model has been extensively used in predictive applications, the analysis of the paper focuses on the extracted loadings and latent features and their corresponding economic interpretation as an alternative approach to understanding the dynamics of price movements in the stock market.

\section{Dynamic Factor Model}
The dynamic factor model (DFM) is another linear model that combines the features of both the PCA and APT model and may be regarded as an unsupervised time series extension of the latter \cite{DFM}. Similar to the APT model, the DFM is a multi-factor model that may be used to explain stock price returns. However, unlike most asset pricing factor models that rely on a pre-defined set of factors, the DFM does not require such inputs. Instead, it estimates the factors directly from the observed data, offering valuable insights into the dynamics of price movements in a data-driven and unsupervised manner similar to PCA. Suppose $R_{it}$ is the return of stock $i$ at time $t$. The model assumes the following dynamics.
\begin{align}
    R_{it} &= \boldsymbol{\beta_i}^\top \boldsymbol{F_t} + \sigma_i Z_{it}, \\
    \boldsymbol{F_t} &= \boldsymbol{\Lambda_1}\boldsymbol{F_{t-1}} + \boldsymbol{\Lambda_2}\boldsymbol{F_{t-2}} + \cdots + \boldsymbol{\Lambda_p}\boldsymbol{F_{t-p}} + \boldsymbol{\varepsilon_t}, \\
    Z_{it} &= \psi_{i1} Z_{i(t-1)} + \psi_{i2} Z_{i(t-2)} + \cdots + \psi_{iq} Z_{i(t-q)} + \gamma_{it},
\end{align}
where $\boldsymbol{F_t}$ is a vector of $n$ common factors at time $t$, $Z_{it}$ is the stock-specific factor of stock $i$ at time $t$, $\boldsymbol{\beta_i}$ is the vector of loadings of stock $i$ for the common factors, $\sigma_i$ is the loading of stock $i$ for the stock-specific factor, $\boldsymbol{\Lambda_j}$ is an $n \times n$ vector autoregressive coefficient matrix for $\boldsymbol{F_{t - j}}$, $\psi_{ij}$ is an autoregressive coefficient for $Z_{i(t - j)}$, and $\boldsymbol{\varepsilon_t} \sim N\left(\boldsymbol{0}, \boldsymbol{I_n}\right)$ and $\gamma_{it} \sim N(0,1)$ are Gaussian noise processes. Furthermore, $\gamma_{it}$ and $\gamma_{jt'}$ are independent for all $i \neq j$ and any $t, t'$. The model thus assumes that the return of stock $i$ at time $t$ is a combination of two components---common and idiosyncratic. The common component is governed by a linear combination of the common factors $\boldsymbol{F_t}$ whereas the idiosyncratic component is governed by $Z_{it}$, a stock-specific factor. 

In addition, the model assumes that stock price returns follow a Gaussian distribution similar to other literature \cite{Kendall1953,Osborne1959,BlackScholes,officer_distribution_1972,roll_empirical_1980,phelan1997probability,marathe_validity_2005,hull2016options,li2023explanationdistributioncharacteristicsstock}. Under the stationarity assumption of the vector autoregressive and autoregressive processes and the model specifications, the common factors $\boldsymbol{F_t} \sim N\left(\boldsymbol{0}, \boldsymbol{\Sigma_F}\right)$ and stock-specific factors $Z_{it} \sim N(0, \sigma_{Zi})$ follow a Gaussian distribution for all $i$ and $t$, where $\boldsymbol{\Sigma_F}$ is some $n \times n$ covariance matrix and $\sigma_{Zi} > 0$. Hence, since $\boldsymbol{F_t}$ and $Z_{it}$ are independent by construction, $R_{it}$ follows a Gaussian distribution with parameters
\begin{equation}
    \mathbb{E}(R_{it}) = \mathbb{E}\left(\boldsymbol{\beta_i}^\top \boldsymbol{F_t}\right) + \mathbb{E}\left(\sigma_i Z_{it}\right) = 0
\end{equation}
and
\begin{equation}
    \text{Var}(R_{it}) = \text{Var}\left(\boldsymbol{\beta_i}^\top \boldsymbol{F_t}\right) + \text{Var}\left(\sigma_i Z_{it}\right) = \boldsymbol{\beta_i}^\top \boldsymbol{\Sigma_F} \boldsymbol{\beta_i} + \sigma_i^2 \sigma_{Zi}^2. \label{eqn:const_var}
\end{equation}

It is worth noting that Eqn. \ref{eqn:const_var} highlights the model's assumption of having constant variance across time. Hence, it is unable to account for conditional heteroskedasticities, such as volatility clustering in periods of high uncertainty. Nonetheless, the DFM remains a versatile and powerful tool with a wide range of applications across various domains. For instance, it has been effectively used in nowcasting economic indicators, demonstrating both efficiency and accuracy in providing timely insights crucial for policymakers, financial analysts, and other stakeholders involved in decision-making processes \cite{luciani_nowcasting_2018,hayashi_nowcasting_2022,chernis2020three}. Beyond nowcasting, the DFM is also extensively applied in the analysis of business cycles, inflation dynamics, and structural analysis \cite{stock1999forecasting,boivin2006has}. In contrast to these predictive applications, the paper focuses on analyzing the extracted loadings, common factors, and their evolution processes along with their economic interpretations as an alternative framework to understand the complex dynamics of stock price movements, complementing established market theories.

\subsection{Model Fitting}
One approach to fit the DFM is to formulate the model as a state-space model and apply Kalman methods and maximum likelihood estimation. The mathematical details of the fitting methodology are presented below.

Suppose $\left\{\boldsymbol{Y_t}\right\}$ is an observed time series process. A linear Gaussian state-space model assumes the following dynamics
\begin{align}
    \boldsymbol{Y_t} &= \boldsymbol{M} \boldsymbol{X_t} + \boldsymbol{\epsilon_t}, \\
    \boldsymbol{X_t} &= \boldsymbol{T} \boldsymbol{X_{t-1}} + \boldsymbol{\eta_t},
\end{align}
where $\left\{\boldsymbol{X_t}\right\}$ is an unobserved latent factor process, $\boldsymbol{M}$ is the measurement loading matrix, $\boldsymbol{T}$ is the transition loading matrix, and $\boldsymbol{\epsilon_t} \sim N\left(\boldsymbol{0}, \boldsymbol{\Sigma_\epsilon}\right), \boldsymbol{\eta_t} \sim N\left(\boldsymbol{0}, \boldsymbol{\Sigma_\eta}\right)$ are Gaussian noise processes.

Denote the following for convenience
\begin{equation}
    \boldsymbol{R_t} \coloneqq 
        \begin{bmatrix}
            R_{1t} \\
            R_{2t} \\
            \vdots \\
            R_{St}
        \end{bmatrix},
    \qquad
    \boldsymbol{\beta} \coloneqq 
        \begin{bmatrix}
            \boldsymbol{\beta_1}^\top \\
            \boldsymbol{\beta_2}^\top \\
            \vdots \\
            \boldsymbol{\beta_S}^\top \\
        \end{bmatrix},
    \qquad
    \boldsymbol{\sigma} \coloneqq 
        \begin{bmatrix}
            \sigma_1 & 0 & \cdots & 0 \\
            0 & \sigma_2 & \cdots & 0 \\
            \vdots & \vdots & \ddots & \vdots \\
            0 & 0 & \cdots & \sigma_S
        \end{bmatrix},
\end{equation}
\begin{equation}
    \boldsymbol{\tilde{Z}_t} \coloneqq 
        \begin{bmatrix}
            \tilde{Z}_{1t} \\
            \tilde{Z}_{2t} \\
            \vdots \\
            \tilde{Z}_{St}
        \end{bmatrix},    
    \qquad
    \boldsymbol{\Psi_j} \coloneqq
        \begin{bmatrix}
            \psi_{1j} & 0 & \cdots & 0 \\
            0 & \psi_{2j} & \cdots & 0 \\
            \vdots & \vdots & \ddots & \vdots \\
            0 & 0 & \cdots & \psi_{Sj}
        \end{bmatrix},
\end{equation}
where $\boldsymbol{R_t}$ is the vector of $S$ stock returns at time $t$, $\boldsymbol{\beta}$ is the combined loadings matrix of all $S$ stocks for the common factors, $\boldsymbol{\sigma}$ is the diagonal loadings matrix of all $S$ stocks for the stock-specific factors, $\boldsymbol{\tilde{Z}_t}$ is the vector of $\tilde{Z}_{it} = \sigma_i Z_{it}$, and $\boldsymbol{\Psi_j}$ is the diagonal matrix containing the $j$th AR coefficients of each $Z_{i(t - j)}$. The DFM can be formulated as a state-space model as follows.

\begin{equation}
    \underbrace{\boldsymbol{R_t}}_{\boldsymbol{Y_t}} = 
        \underbrace{\begin{bmatrix}
            \boldsymbol{\beta} & \boldsymbol{0} & \cdots & \boldsymbol{0} & \boldsymbol{I_s} & \boldsymbol{0} & \cdots & \boldsymbol{0} 
        \end{bmatrix}}_{\boldsymbol{M}}
        \underbrace{\begin{bmatrix}
            \boldsymbol{F_t} \\
            \boldsymbol{F_{t-1}} \\
            \vdots \\
            \boldsymbol{F_{t-p+1}} \\
            \boldsymbol{\tilde{Z}_t} \\
            \boldsymbol{\tilde{Z}_{t-1}} \\
            \vdots \\
            \boldsymbol{\tilde{Z}_{t-q+1}}
        \end{bmatrix}}_{\boldsymbol{X_t}}
        + ~\boldsymbol{\epsilon_t},
\end{equation}
\begin{equation}
    \underbrace{\begin{bmatrix}
        \boldsymbol{F_t} \\
        \vdots \\
        \boldsymbol{F_{t-p+2}} \\
        \boldsymbol{F_{t-p+1}} \\
        \boldsymbol{\tilde{Z}_t} \\
        \vdots \\
        \boldsymbol{\tilde{Z}_{t-q+2}} \\
        \boldsymbol{\tilde{Z}_{t-q+1}}
    \end{bmatrix}}_{\boldsymbol{X_t}} =
    \underbrace{\begin{bmatrix}
        \boldsymbol{\Lambda_1} & \cdots & \boldsymbol{\Lambda_{p-1}} & \boldsymbol{\Lambda_p} & \boldsymbol{0} & \cdots & \boldsymbol{0} & \boldsymbol{0} \\
        \boldsymbol{I_n} & \cdots & \boldsymbol{0} & \boldsymbol{0} & \boldsymbol{0} & \cdots & \boldsymbol{0} & \boldsymbol{0} \\
        \vdots & \ddots & \vdots & \vdots & \vdots & \ddots & \vdots & \vdots \\
        \boldsymbol{0} & \cdots & \boldsymbol{I_n} & \boldsymbol{0} & \boldsymbol{0} & \cdots & \boldsymbol{0} & \boldsymbol{0} \\
        \boldsymbol{0} & \cdots & \boldsymbol{0} & \boldsymbol{0} & \boldsymbol{\Psi_1} & \cdots & \boldsymbol{\Psi_{q-1}} & \boldsymbol{\Psi_{q}} \\
        \boldsymbol{0} & \cdots & \boldsymbol{0} & \boldsymbol{0} & \boldsymbol{I_s} & \cdots & \boldsymbol{0} & \boldsymbol{0} \\
        \vdots & \ddots & \vdots & \vdots & \vdots & \ddots & \vdots & \vdots \\
        \boldsymbol{0} & \cdots & \boldsymbol{0} & \boldsymbol{0} & \boldsymbol{0} & \cdots & \boldsymbol{I_s} & \boldsymbol{0} \\
    \end{bmatrix}}_{\boldsymbol{T}}
    \underbrace{\begin{bmatrix}
        \boldsymbol{F_{t-1}} \\
        \vdots \\
        \boldsymbol{F_{t-p+1}} \\
        \boldsymbol{F_{t-p}} \\
        \boldsymbol{\tilde{Z}_{t-1}} \\
        \vdots \\
        \boldsymbol{\tilde{Z}_{t-q+1}} \\
        \boldsymbol{\tilde{Z}_{t-q}}
    \end{bmatrix}}_{\boldsymbol{X_{t-1}}}
    + ~ \boldsymbol{\eta_t},
\end{equation}
with $\boldsymbol{\Sigma_\epsilon} = \boldsymbol{0}$ and $\boldsymbol{\Sigma_\eta} = \text{diag}\left(\boldsymbol{I_n}, \boldsymbol{0}, \ldots, \boldsymbol{0}, \boldsymbol{\sigma}^2, \boldsymbol{0}, \ldots, \boldsymbol{0}\right)$.

For a fixed set of parameters $\boldsymbol{M}, \boldsymbol{T}$, $\boldsymbol{\Sigma_\epsilon}$, and $\boldsymbol{\Sigma_\eta}$, the Kalman filter may be used to estimate the state of the model $\boldsymbol{X_t}$, and consequently the common factors $\boldsymbol{F_t}$ and stock-specific factors $Z_{it}$. For convenience, denote
\begin{align}
    \boldsymbol{Y_{1:t}} &\coloneqq \left\{\boldsymbol{Y_1}, \boldsymbol{Y_2}, \ldots, \boldsymbol{Y_t}\right\}, \\
    \boldsymbol{\mu_{t \mid t'}} &\coloneqq \mathbb{E}\left(\boldsymbol{X_t} \mid \boldsymbol{Y_{1:t'}}\right), \\
    \boldsymbol{\Sigma_{t \mid t'}} &\coloneqq \text{Cov}\left(\boldsymbol{X_t} \mid \boldsymbol{Y_{1:t'}}\right).
\end{align}
The Kalman filter prediction step predicts the current state of the system as
$\boldsymbol{X_t} \mid \boldsymbol{Y_{1:t - 1}} \sim N\left(\boldsymbol{\mu_{t \mid t - 1}}, \boldsymbol{\Sigma_{t \mid t - 1}}\right)$, where 
\begin{align}
    \boldsymbol{\mu_{t \mid t - 1}} &= \boldsymbol{T}\boldsymbol{\mu_{t - 1 \mid t - 1}}, \\
    \boldsymbol{\Sigma_{t \mid t - 1}} &= \boldsymbol{T}\boldsymbol{\Sigma_{t - 1 \mid t - 1}}\boldsymbol{T}^\top + \boldsymbol{\Sigma_\eta}.
\end{align}
The Kalman filter update step then combines knowledge about the predicted state $\boldsymbol{X_t}$ and the new observation $\boldsymbol{Y_t}$ to produce an updated estimate of the current state of the system as $\boldsymbol{X_t} \mid \boldsymbol{Y_{1:t}} \sim N\left(\boldsymbol{\mu_{t \mid t}}, \boldsymbol{\Sigma_{t \mid t}}\right)$, where
\begin{align}
    \boldsymbol{K_t} \coloneqq& \boldsymbol{\Sigma_{t \mid t - 1}} \boldsymbol{M}^\top \left(\boldsymbol{M}\boldsymbol{\Sigma_{t \mid t - 1}} \boldsymbol{M}^\top + \boldsymbol{\Sigma_\epsilon}\right)^{-1}, \\
    \boldsymbol{\mu_{t \mid t}} =& \boldsymbol{\mu_{t \mid t - 1}} + \boldsymbol{K_t} \left(\boldsymbol{Y_t} - \boldsymbol{M} \boldsymbol{\mu_{t \mid t - 1}}\right), \\
    \boldsymbol{\Sigma_{t \mid t}} =& \boldsymbol{\Sigma_{t \mid t - 1}} - \boldsymbol{K_t} \boldsymbol{M}\boldsymbol{\Sigma_{t \mid t - 1}}.
\end{align}
Hence, the Kalman filter uses past and current observations to estimate the current state of the system. The optimal parameters may then be obtained via maximum likelihood estimation.

Additionally, Kalman smoothing may be used at time $t' > t$ to combine knowledge about all observations until time $t'$ to produce an updated estimate of the state of the system at time $t$ as $\boldsymbol{X_t} \mid \boldsymbol{Y_{1:t'}} \sim N\left(\boldsymbol{\mu_{t \mid t'}}, \boldsymbol{\Sigma_{t \mid t'}}\right)$, where
\begin{align}
    \boldsymbol{J_t} \coloneqq& \boldsymbol{\Sigma_{t \mid t}} \boldsymbol{T}^\top \boldsymbol{\Sigma_{t + 1 \mid t}}^{-1}, \\
    \boldsymbol{\mu_{t \mid t'}} =& \boldsymbol{\mu_{t \mid t}} + \boldsymbol{J_t} \left(\boldsymbol{\mu_{t + 1 \mid t'}} - \boldsymbol{\mu_{t + 1 \mid t}}\right), \\
    \boldsymbol{\Sigma_{t \mid t'}} =& \boldsymbol{\Sigma_{t \mid t}} + \boldsymbol{J_t} \left(\boldsymbol{\Sigma_{t + 1 \mid t'}} - \boldsymbol{\Sigma_{t + 1 \mid t}}\right) \boldsymbol{J_t}^\top.
\end{align}
Unlike the Kalman filter, Kalman smoothing uses past, current, and future observations to estimate the current state of the system.

\subsection{Model Validation}
Meanwhile, the theoretical and empirical validity of the DFM relies on specifying the correct number of common factors $n$. While previous research often set $n$ based on prior knowledge and existing studies, \cite{bai2002determining} provided three information criteria as statistical measures to consistently estimate $n$ from a given dataset. These information criteria, extensively used in literature \cite{
bai2003,stock2022,STOCK2016415,giglio2022}, are expressed as
\begin{align}
    \text{IC}_1(n) &= \ln V(n) + n \left(\dfrac{N + T}{NT}\right) \ln \left(\dfrac{NT}{N + T}\right), \label{eqn:ic1} \\
    \text{IC}_2(n) &= \ln V(n) + n \left(\dfrac{N + T}{NT}\right) \ln \min\left\{N, T\right\}, \label{eqn:ic2} \\
    \text{IC}_3(n) &= \ln V(n) + n \left(\dfrac{\ln \min\left\{N, T\right\}}{\min\left\{N, T\right\}}\right), \label{eqn:ic3}
\end{align}
where $V(n)$ is the mean of the squared residuals when $n$ common factors are estimated via PCA. In contrast, \cite{Onatski2009} proposed a hypothesis testing procedure for determining the number of common factors $n$. More recently, \cite{MOLEROGONZALEZ2023103816} also provided an alternative method based on Random Matrix Theory (RMT). 

In addition, following the approach in \cite{MOLEROGONZALEZ2023103816}, additional analyses can be performed to assess the model's alignment with established market theories. Examining the relationship between the common factors $\boldsymbol{F_t}$ and the composite index of a market may reveal the model's ability to capture systematic market movements and distinguish idiosyncratic components of stock price dynamics. Additionally, analyzing the correlation between the factor loadings $\boldsymbol{\beta_i}$ and the CAPM beta, as defined in Eqn.~\ref{eqn:capm}, offers insights into how well the model reflects a conventional measure of exposure to systematic risks. By selecting an appropriate number of factors $n$ with corresponding loadings and common factors that align with market theories, the empirical validity of the DFM can be effectively demonstrated.

\subsection{Model Implementation}
The fitting and validation procedures for the DFM described above are implemented in the \texttt{DynamicFactorAnalysis} Python package accessible at \url{https://github.com/briangodwinlim/DynamicFactorAnalysis}. Other common data science libraries are also used in the implementation.

\section{Results}
To reiterate, the paper analyzes the extracted loadings $\boldsymbol{\beta_i}$ and common factors $\boldsymbol{F_t}$ to provide an alternative perspective on the dynamics of stock price movements, distinguishing itself from recent developments in literature, particularly \cite{MOLEROGONZALEZ2023103816}, which did not provide a subsequent investigation into $\boldsymbol{\beta_i}$ and $\boldsymbol{F_t}$ after using RMT to identify the number of common factors underlying stock price dynamics. Whereas \cite{MOLEROGONZALEZ2023103816} primarily focused on factor dimensionality, the results of this paper specifically analyzed and interpreted $\boldsymbol{\beta_i}$ and $\boldsymbol{F_t}$ in relation to the broader framework of known econometric and market facts.

To this end, the Philippine Stock Exchange (PSE) is considered due to its distinct economic landscape and investor behavior. This approach not only broadens insights into stock price dynamics in a unique economic context but also demonstrates the robustness of the DFM in providing insights aligned with established market theories. The historical stock price data used in the DFM is obtained using the Python library \texttt{fastquant} which wraps the process for requesting data using the Phisix API\footnote{\url{https://github.com/phisix-org/phisix}}. In addition, the data for the Philippine Stock Exchange Index (PSEi), the composite index of the PSE, is obtained from \url{https://stooq.com/q/d/?s=\%5Epsei}.

The period from January 1, 2015, to December 31, 2020, is considered. Following the data cleaning procedure of existing works \cite{neszveda2025aspiration,feder2024global,feng2019temporal} and to maintain the integrity of the dataset, stocks with more than 1\% missing observations are dropped from the analysis. This results in a dataset containing 72 stocks, more than the top 30 in terms of market capitalization included in the PSEi. Given the objective of the paper, the above procedure would result in the exclusion of generally low-volume stocks that would have a comparatively smaller contribution to overall market dynamics. The high-volume stocks remaining in the analysis would have more weight, guaranteeing generalizability even with this data cleaning procedure.

It is also worth mentioning that while the inclusion of the coronavirus pandemic period brings significant volatility to the data, an important aspect of the present paper is to demonstrate the ability of the model to provide robust insights into the market even in the presence of extreme events. The inclusion of the pandemic period thus provides a rigorous test case to evaluate the robustness of the model under such unprecedented conditions, in line with \cite{MOLEROGONZALEZ2023103816}.

\begin{figure}[h!]
    \centering
    \includegraphics[scale=0.50]{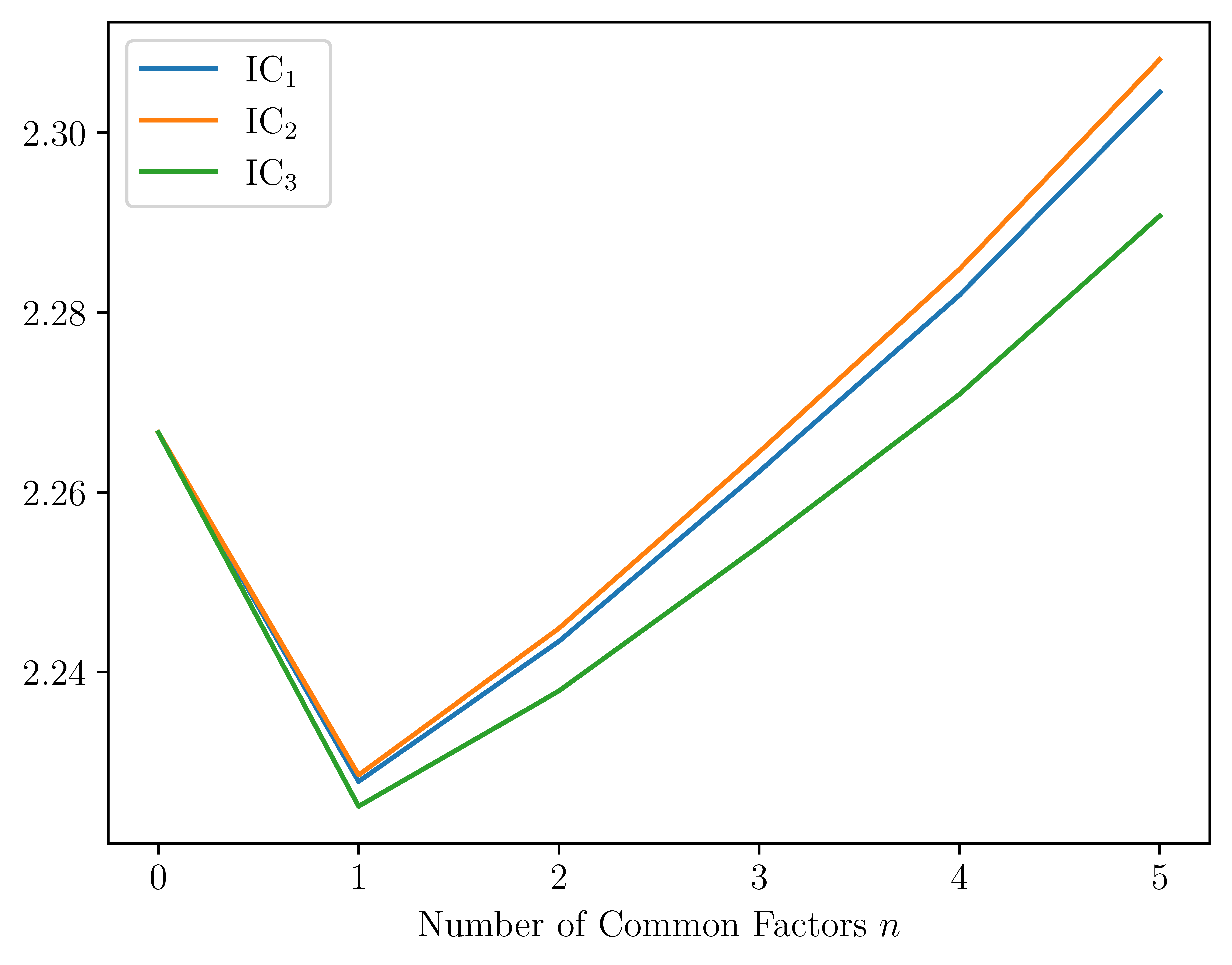}
    \caption{Information criteria for different number of common factors $n$.}
    \label{fig:ic}
\end{figure}

To ensure stationarity of the data, the percentage return $R_{it}$ is considered. If $S_{it}$ is the closing price of stock $i$ at time $t$, the percentage return is obtained as
\begin{equation}
    R_{it} = \dfrac{S_{it} - S_{i(t-1)}}{S_{i(t-1)}}.
\end{equation}
The information criteria in Eqns.~\ref{eqn:ic1}, \ref{eqn:ic2}, and \ref{eqn:ic3} across different values of $n$ are then calculated and presented in Fig.~\ref{fig:ic}. The figure shows that the information criteria are lower for models with $n = 1$ or $n = 2$ factors, specifically when compared with a white noise model containing zero factors. This validates the choice of a factor model as it suggests the inclusion of at least one factor considerably improves fit with the data. While an analysis based entirely on the information criteria would suggest that a model with $n = 1$ common factor fits the data best, a model with $n = 2$ common factors also exhibits a comparably close fit, making it a viable alternative for consideration. 

\subsection{One-Factor Model}
\begin{figure}[h!]
    \centering
    \includegraphics[scale=0.45]{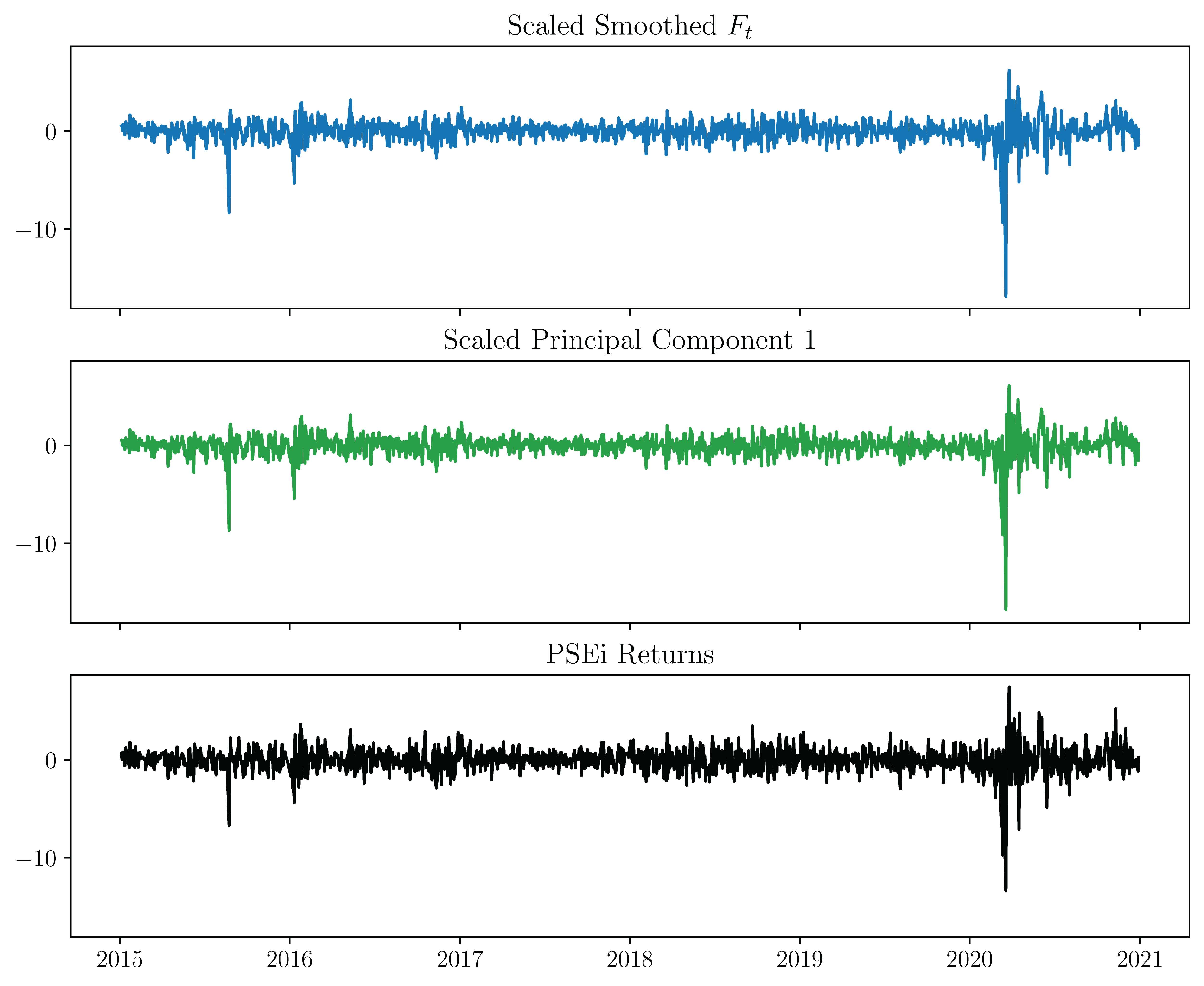}
    \caption{The common factor $F_t$ for DFM ($n = 1$, $p = 3$, $q = 5$), the first principal component, and the PSEi return from 2015 to 2020.}
    \label{fig:res135}
\end{figure}

First, the model with $n = 1$ common factor following an AR(3) process and stock-specific factors following AR(5) processes is considered, where $p$ and $q$ are chosen based on the Bayesian Information Criterion (BIC). As a validation, the common factor $F_t$ is compared against the return of PSEi as seen in Fig.~\ref{fig:res135}. Generally, composite indices contain a diversified portfolio of stocks in a particular market. This diversification removes stock-related movements and risks leaving only systematic movements and risks. Consequently, the PSEi is commonly used as a proxy for systematic movements in the market. The (Kalman smoothed) common factor $F_t$ and the PSEi returns exhibit a correlation of 0.9283, establishing the former as a viable indicator for systematic movements in the market. Nevertheless, in contrast to the PSEi, which only covers the top 30 stocks with the largest market capitalization, the common factor $F_t$ captures the systematic movements across the broader market, reflecting the overall stock price movements. This accounts for the discrepancy between the common factor and the PSEi, and why their correlation coefficient falls short of perfect association. 

Nonetheless, the common factor $F_t$ is still versatile in capturing market conditions at various points in time. Notably, it accurately reflects significant market events such as the sharp downturn on August 24, 2015, resulting from the global financial market sell-offs due to concerns about China's economy. It also reflects the high economic volatility during the first two quarters of 2020 when the Philippine economy underwent lockdown due to the novel coronavirus disease. The common factor remains informative in reflecting the challenges during the subsequent recovery process. Other similar observations can also be noted such as volatility clustering during times of high economic uncertainty. This further supports the common factor $F_t$ representing systematic movements in the market.

The common factor $F_t$ is also estimated to evolve according to 
\begin{equation}
    F_t = 0.1256 F_{t-1} + 0.0225 F_{t-2} + 0.1380 F_{t-3} + \varepsilon_t,
\end{equation}
where only the parameter $\Lambda_2 = 0.0225$ is not statistically significant. This suggests that systematic shocks in the market today are expected to persist for at least 3 trading days. Such persistence may reveal important aspects of market behavior. For example, policy-related shocks might influence the market over an extended period, indicating the need for caution when announcing or implementing policies that are expected to impact the market \cite{li2010impact,chatziantoniou2013stock}.

Moreover, PCA is also investigated as a comparison. Table~\ref{tab:corr135} presents a summary of the correlation figures. The (Kalman smoothed) common factor $F_t$ exhibits a correlation of 0.9975 with the first principal component as illustrated in Fig.~\ref{fig:res135}. This principal component also has a strong correlation of 0.9144 with the PSEi returns. While these results might suggest that the DFM performs similarly to PCA in explaining systematic stock price movements, it must be noted that the two models are fundamentally distinct. PCA derives principal components as a linear combination of stock price returns, while the DFM explains stock price returns as a linear combination of some underlying common factors. The latter is also a time series model, accounting for the dynamics of stock price returns across time, while the former is not. Moreover, when considering $n > 1$ factors, the first principal component will remain constant whereas the common factors $\boldsymbol{F_t}$ will adapt depending on the model specifications. Hence, the DFM can capture the dynamics of stock price movements better than PCA. 

\begin{table*}[p]
    \centering
    \caption{Summary of correlation figures for DFM ($n = 1, p = 3, q = 5$).}
    \label{tab:corr135}
    \setlength{\tabcolsep}{0.6em}
    \renewcommand{\arraystretch}{1.25}
    \scalebox{0.9}{
    \begin{tabular}{|cc|c|}
        \hline
        Series 1 & Series 2 & Correlation \\
        \hline
        CAPM $\beta_i$ & $\beta_i$ & 0.8348 \\
        $F_t$ & PSEi & 0.9283 \\
        $F_t$ & PC1 & 0.9975 \\
        PC1 & PSEi & 0.9144 \\
        \hline
    \end{tabular}
    }
\end{table*}

\begin{figure}[h!]
    \centering
    \includegraphics[scale=0.45]{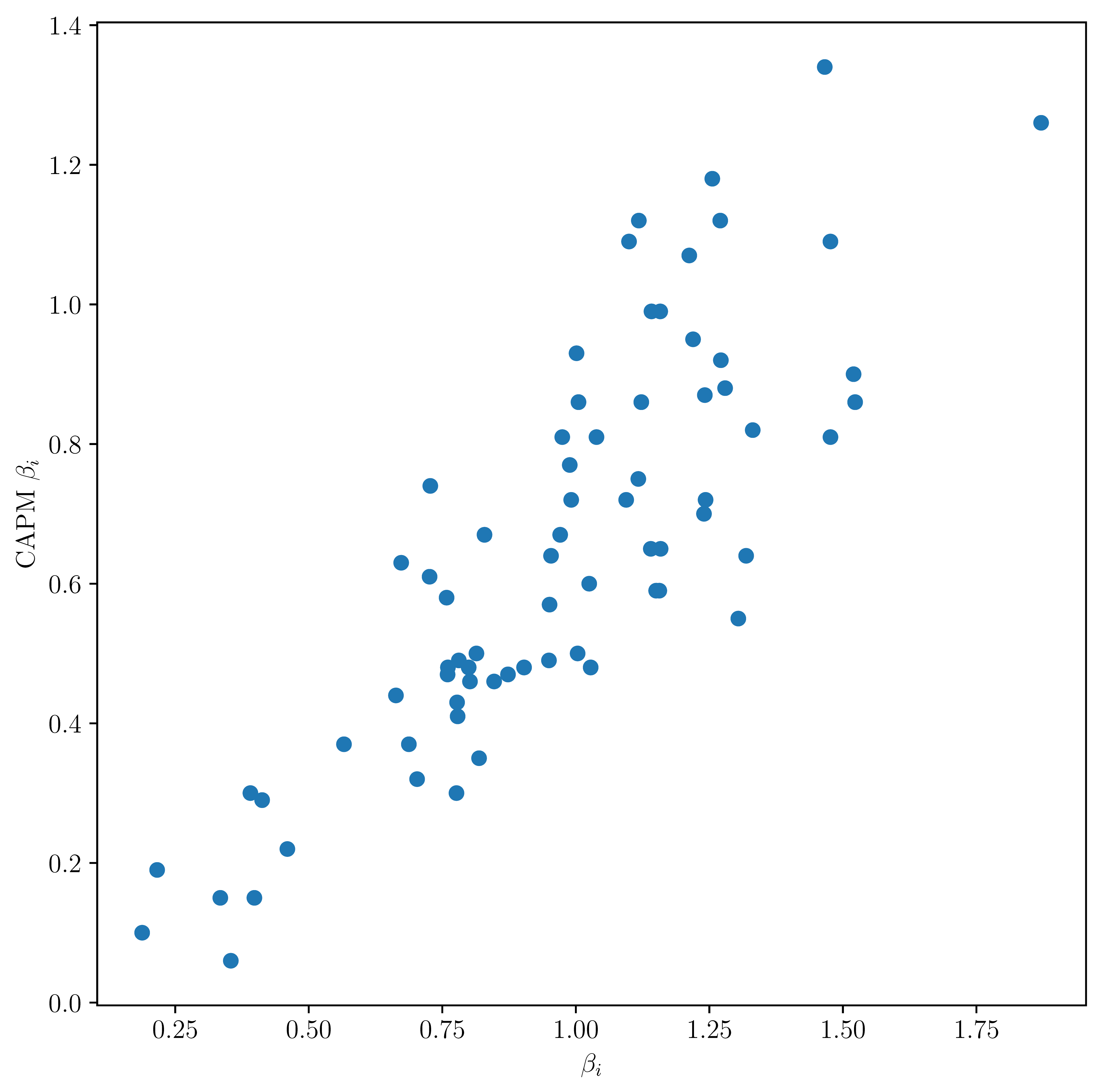}
    \caption{Scatterplot of the loading $\beta_{i}$ for DFM ($n = 1$, $p = 3$, $q = 5$) and the CAPM beta.}
    \label{fig:beta135}
\end{figure}

As further model validation, the loading $\beta_{i}$ is compared against its corresponding CAPM beta, widely accepted as a measure of exposure to market risks. Fig.~\ref{fig:beta135} presents the scatterplot between the loading $\beta_{i}$ and the CAPM beta\footnote{\url{https://www.barrons.com/market-data/stocks}}. The figure shows a strong positive relationship between the two measures with a correlation of 0.8348, validating the relationship of $\beta_{i}$ and exposure to systematic movements. Table~\ref{tab:res135} then presents the summary of the loading $\beta_{i}$ and $\sigma_i$ for the 72 stocks included in the analysis. At a significance level of $\alpha = 0.05$, the loading $\beta_i$ for the common factor $F_t$ is statistically significant for all but one stock\footnote{The loading $\beta_i$ for stock LC is not statistically significant due to a significant stock-specific shock---the sizable closure order against its operations \cite{denr,catajanPhilippineMinesContinue2021}---that lasted from 2017 to 2020. This is expected to have a pronounced impact on its financial outlook, thereby overshadowing general market conditions.}. This result highlights that the common factor $F_t$ explains price movements in the market, capturing the systematic or market dynamics similar to the PSEi. To illustrate, more mature and developed stocks such as AC, GTCAP, JFC, SM, and URC have relatively balanced $\beta_{i}$ and $\sigma_i$ values. In contrast, less mature or more volatile stocks such as ABA, BRN, FNI, PXP, and SSP have relatively higher $\sigma_i$ values, indicating a larger contribution from idiosyncratic movements due to stock-specific shocks. Hence, the ratio between $\beta_i$ and $\sigma_i$ may also hint at the relative sensitivity of a stock to market-specific or stock-specific volatilities.

\begin{table*}[p]
    \centering
    \caption{Loadings for DFM ($n = 1$, $p = 3$, $q = 5$).}
    \label{tab:res135}
    \setlength{\tabcolsep}{0.6em}
    \begin{minipage}{0.5\linewidth}
    \centering
    \scalebox{0.9}{
    \begin{tabular}{|c|cc|}
        \hline
        Stock & $\beta_{i}$ & $\sigma_{i}$ \\
        \hline
        2GO & 0.8470 & 3.9217 \\
        ABA & 0.9707 & 3.2377 \\
        AC & 1.2704 & 1.4373 \\
        AEV & 1.0996 & 1.8298 \\
        AGI & 1.2715 & 1.7219 \\
        ALI & 1.8711 & 1.6140 \\
        AP & 0.7276 & 1.5130 \\
        BDO & 1.2555 & 1.4833 \\
        BEL & 0.8189 & 1.6531 \\
        BLOOM & 1.5229 & 2.7239 \\
        BPI & 1.0013 & 1.4661 \\
        BRN & 1.2399 & 3.3060 \\
        CEB & 1.3312 & 2.4174 \\
        CHIB & 0.4127 & 0.9347 \\
        CNPF & 0.6633 & 1.7277 \\
        COSCO & 0.7777 & 1.4535 \\
        CPG & 0.8730 & 2.2062 \\
        DD & 0.7787 & 2.4391 \\
        DMC & 1.2795 & 1.8911 \\
        DNL & 1.1504 & 2.2307 \\
        EEI & 0.8018 & 2.1847 \\
        EW & 1.0251 & 1.8732 \\
        FGEN & 0.8139 & 1.8883 \\
        FLI & 1.0034 & 1.5730 \\
        FNI & 1.2429 & 6.0448 \\
        FPH & 0.7029 & 1.2352 \\
        GERI & 1.0278 & 2.1231 \\
        GLO & 0.7262 & 1.8556 \\
        GMA7 & 0.3908 & 1.3900 \\
        GTCAP & 1.2195 & 1.8579 \\
        HOUSE & 0.3541 & 1.8103 \\
        ICT & 1.1416 & 1.8282 \\
        IMI & 0.9508 & 2.4919 \\
        JFC & 1.1580 & 1.7656 \\
        JGS & 1.4660 & 1.8076 \\
        LC & 0.1881$^\ns$ & 3.0327 \\
        \hline
    \end{tabular}
    }
    \end{minipage}\hfill
    \begin{minipage}{0.5\linewidth}
    \centering
    \scalebox{0.9}{
    \begin{tabular}{|c|cc|}
        \hline
        Stock & $\beta_{i}$ & $\sigma_{i}$ \\
        \hline
        LPZ & 0.7766 & 2.0411 \\
        LTG & 0.9536 & 2.1000 \\
        MAXS & 1.1587 & 2.0482 \\
        MBT & 1.1226 & 1.5110 \\
        MEG & 1.5200 & 1.7571 \\
        MER & 0.6730 & 1.5231 \\
        MPI & 1.2415 & 1.9402 \\
        MWC & 0.7810 & 2.2912 \\
        MWIDE & 0.9886 & 2.3709 \\
        NI & 0.3982 & 2.5281 \\
        NIKL & 1.3189 & 3.6951 \\
        PCOR & 0.9497 & 2.1039 \\
        PGOLD & 0.7605 & 1.6123 \\
        PLC & 1.1563 & 2.4863 \\
        PNB & 0.7994 & 1.5993 \\
        PNX & 0.4600 & 1.9619 \\
        PX & 0.5659 & 2.4845 \\
        PXP & 1.3042 & 5.2152 \\
        RLC & 1.4767 & 1.9403 \\
        RRHI & 0.7599 & 1.7353 \\
        SCC & 0.9913 & 2.8519 \\
        SECB & 1.0943 & 1.7228 \\
        SLI & 0.3344 & 2.0121 \\
        SM & 1.1180 & 1.6795 \\
        SMC & 0.7581 & 1.7769 \\
        SMPH & 1.2124 & 1.5934 \\
        SSI & 1.4767 & 2.7815 \\
        SSP & 0.9031 & 3.6809 \\
        STI & 0.6874 & 2.6140 \\
        TECH & 0.9746 & 3.7055 \\
        TEL & 0.8290 & 1.9001 \\
        UBP & 0.2162 & 1.0130 \\
        URC & 1.0050 & 1.7758 \\
        VITA & 1.1404 & 3.6114 \\
        VLL & 1.1170 & 2.1362 \\
        WEB & 1.0387 & 4.8616 \\
        \hline
    \end{tabular}
    }
    \end{minipage}    
\end{table*}

\subsection{Two-Factor Model}
\begin{figure}[h!]
    \centering
    \includegraphics[scale=0.45]{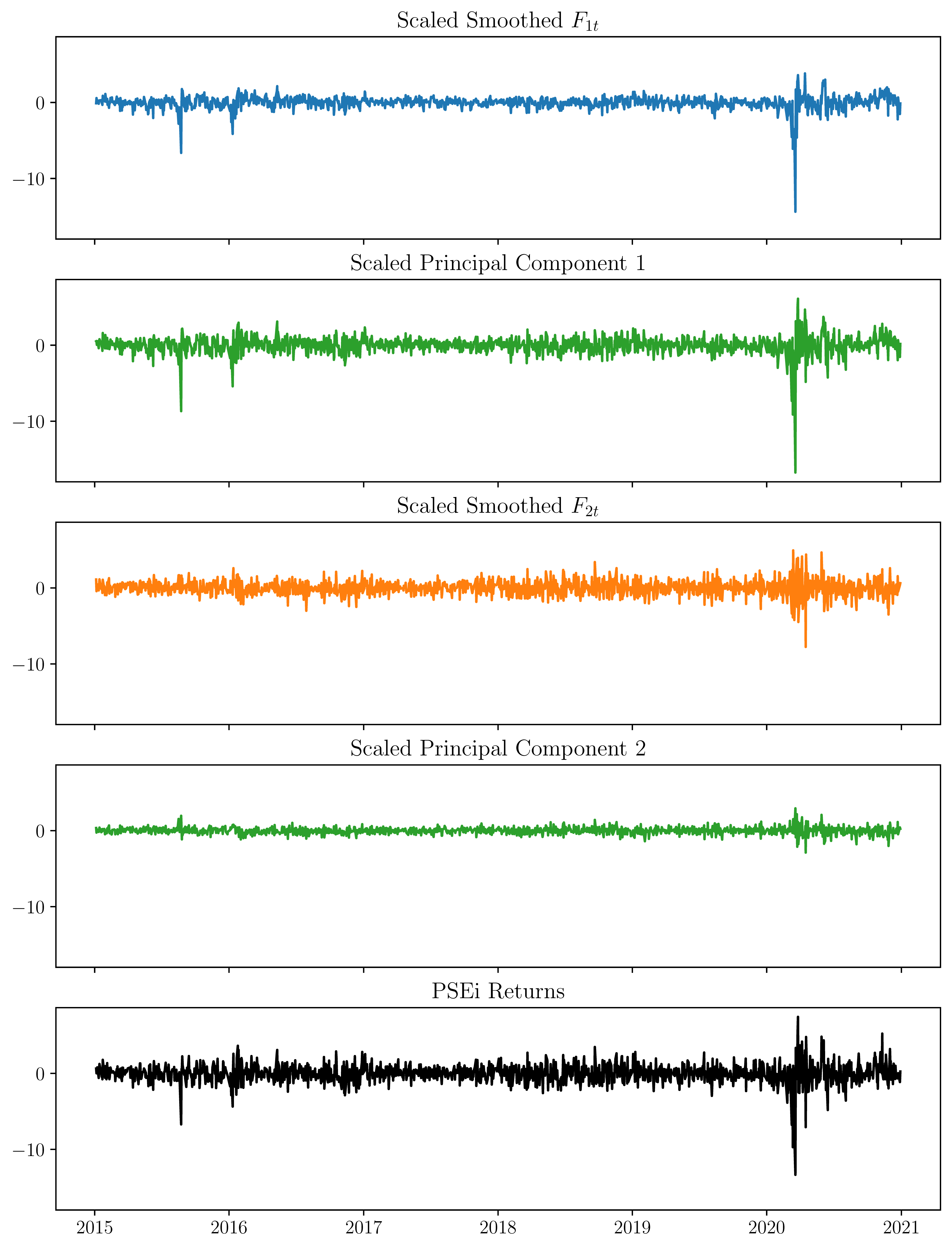}
    \caption{The common factors $\boldsymbol{F_t}$ for DFM ($n = 2$, $p = 2$, $q = 5$), the first two principal components, and the PSEi return from 2015 to 2020.}
    \label{fig:res225}
\end{figure}

Next, the model with $n = 2$ common factors following a VAR(2) process and stock-specific factors following AR(5) processes is considered, where $p$ and $q$ are chosen based on the BIC. As validation, the (Kalman smoothed) common factors $\boldsymbol{F_t}$ are also compared against the PSEi returns and the principal components in Fig.~\ref{fig:res225}. The correlation figures are presented in Table~\ref{tab:corr225}. The first common factor $F_{1t}$ has a correlation of 0.6727 and 0.8982 with the PSEi returns and the first principal component, respectively. The second common factor $F_{2t}$ has a correlation of 0.7704 and 0.8229 with the PSEi returns and the second principal component, respectively. The first and second principal components also have a correlation of 0.9144 and 0.3498 with the PSEi returns, respectively. To reiterate, the common factors $\boldsymbol{F_t}$ of the DFM adapt based on the model specification while the principal components of PCA remain static regardless of the number of factors considered. This may be observed from the common factor $F_{1t}$ in Fig.~\ref{fig:res225} that deviates from the common factor $F_t$ in Fig.~\ref{fig:res135} but still maintains the overall trend. This highlights the distinctive advantage of the DFM over PCA in determining the systematic movements in the PSE. PCA requires the principal components to maximize variance while being mutually orthogonal. The DFM does not explicitly impose such restrictions, thereby allowing more flexibility. 

\begin{table*}[p]
    \centering
    \caption{Summary of correlation figures for DFM ($n = 2, p = 2, q = 5$).}
    \label{tab:corr225}
    \setlength{\tabcolsep}{0.6em}
    \renewcommand{\arraystretch}{1.25}
    \scalebox{0.9}{
    \begin{tabular}{|cc|c|}
        \hline
        Series 1 & Series 2 & Correlation \\
        \hline
        CAPM $\beta_i$ & $\beta_{1i}$ & 0.5132 \\
        $F_{1t}$ & PSEi & 0.6727 \\
        $F_{1t}$ & PC1 & 0.8982 \\
        PC1 & PSEi & 0.9144 \\
        \hline
    \end{tabular}
    \hspace{20pt}
    \begin{tabular}{|cc|c|}
        \hline
        Series 1 & Series 2 & Correlation \\
        \hline
        CAPM $\beta_i$ & $\beta_{2i}$ & 0.8528 \\
        $F_{2t}$ & PSEi & 0.7704 \\
        $F_{2t}$ & PC2 & 0.8229 \\
        PC2 & PSEi & 0.3498 \\
        \hline
    \end{tabular}
    }
\end{table*}

Exploring the common factors in greater detail, it is worth noting that $F_{1t}$ and $F_{2t}$ have a correlation of 0.0798, nearly orthogonal. A linear regression of $F_t$ on $F_{1t}$ and $F_{2t}$ also produced an $R^2$ of 0.998. These results indicate that the model further decomposed $F_t$ into two nearly uncorrelated signals $F_{1t}$ and $F_{2t}$. A visual inspection of Fig.~\ref{fig:res225} reveals that $F_{1t}$ may represent the broader market trend while $F_{2t}$ may represent market uncertainties independent of the general market direction. This also allows for the following interpretation: $\beta_{1i}$ as exposure to market trend, $\beta_{2i}$ as exposure to market volatility, and $\sigma_i$ as exposure to stock-specific volatility. Consequently, this provides a new perspective on portfolio risk management beyond the traditional CAPM framework as investors may now consider two dimensions to market movements.

The common factors $\boldsymbol{F_t} = \left[F_{1t}, F_{2t}\right]^\top$ are then estimated to evolve according to 
\begin{equation}
    \boldsymbol{F_t} =
    \begin{bmatrix}
        \phantom{-}0.4894 & \phantom{-}0.1680 \\
        -0.0262 & -0.2526 \\
    \end{bmatrix} \boldsymbol{F_{t-1}} + 
    \begin{bmatrix}
        \phantom{-}0.5094 & -0.1726 \\
        -0.1627 & -0.0931 \\
    \end{bmatrix} \boldsymbol{F_{t-2}} + 
    \boldsymbol{\varepsilon_t},
\end{equation}
where only the coefficients $-0.0262$ and $-0.0931$ are not statistically significant. Similarly, the results also suggest the persistence of systematic shocks in the market \cite{gil2023persistence}.

\begin{figure}[h!]
    \centering
    \includegraphics[scale=0.45]{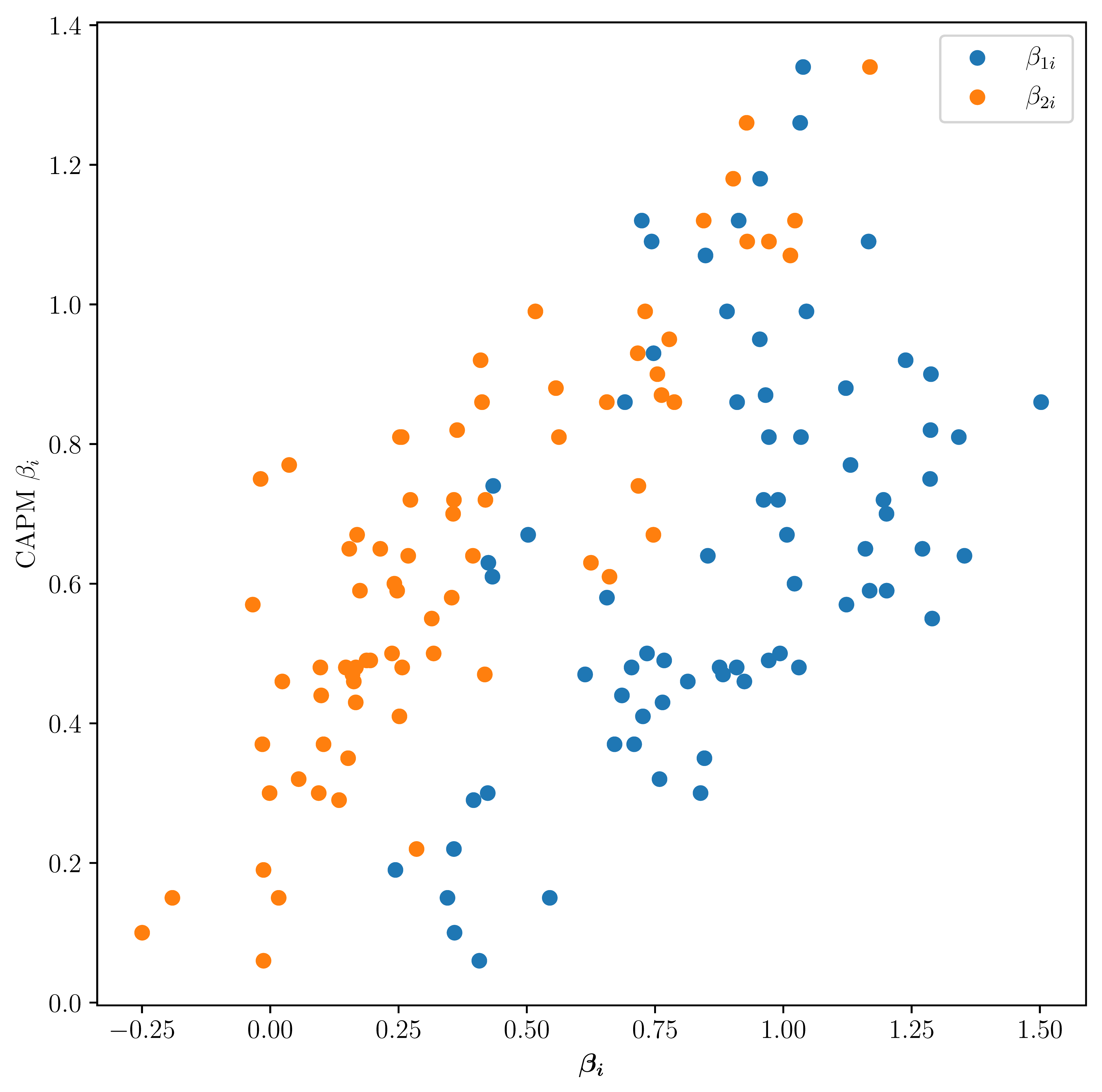}
    \caption{Scatterplot of the loadings $\boldsymbol{\beta_{i}}$ for DFM ($n = 2$, $p = 2$, $q = 5$) and the CAPM beta.}
    \label{fig:beta225}
\end{figure}

As added validation, the loadings $\beta_{1i}$ and $\beta_{2i}$ are also compared against the CAPM beta as presented in Fig.~\ref{fig:beta225}. The CAPM beta exhibits a correlation of 0.5132 and 0.8528 with $\beta_{1i}$ and $\beta_{2i}$, respectively. The lower correlation with $\beta_{1i}$ suggests that the first common factor $F_{1t}$ captures market dynamics not considered by the CAPM. Meanwhile, the higher correlation between $\beta_{2i}$ and the CAPM beta aligns with the interpretation of the second common factor $F_{2t}$ capturing market volatility, a key aspect that CAPM also emphasizes. Despite these differences, both common factors $F_{1t}$ and $F_{2t}$ significantly contribute to explaining systematic stock price movements, highlighting the ability of the model to capture both market trends and more specific sources of volatility. Table~\ref{tab:res225} then presents the summary of the loadings $\beta_{1i}$, $\beta_{2i}$, and $\sigma_i$. The $\beta_{1i}$'s are statistically significant for all stocks while half of the $\beta_{2i}$'s are statistically significant at $\alpha = 0.05$. This suggests that the two-factor model captures additional variance in the data, highlighting how the common factors obtained are viable indicators for systematic movements in the market. 

\begin{table*}[p]
    \centering
    \caption{Loadings for DFM ($n = 2$, $p = 2$, $q = 5$).}
    \label{tab:res225}
    \setlength{\tabcolsep}{0.6em}
    \begin{minipage}{0.5\linewidth}
    \centering
    \scalebox{0.9}{
    \begin{tabular}{|c|ccc|}
        \hline
        Stock & $\beta_{1i}$ & $\beta_{2i}$ & $\sigma_{i}$ \\
        \hline
        2GO & 0.9241 & 0.0236$^\ns$ & 3.8987 \\
        ABA & 1.0070 & 0.1693$^\ns$ & 3.2195 \\
        AC & 0.9129 & 0.8448 & 1.3988 \\
        AEV & 0.7433 & 0.9720 & 1.7430 \\
        AGI & 1.2384 & 0.4100 & 1.7036 \\
        ALI & 1.0328 & 0.9285 & 1.5250 \\
        AP & 0.4345 & 0.7176 & 1.4191 \\
        BDO & 0.9548 & 0.9023 & 1.4394 \\
        BEL & 0.8462 & 0.1517 & 1.6286 \\
        BLOOM & 1.5024 & 0.4127 & 2.6979 \\
        BPI & 0.7470 & 0.7162 & 1.4348 \\
        BRN & 1.2011 & 0.3564 & 3.2954 \\
        CEB & 1.2868 & 0.3642 & 2.3957 \\
        CHIB & 0.3963 & 0.1341 & 0.9321 \\
        CNPF & 0.6854 & 0.0993$^\ns$ & 1.7111 \\
        COSCO & 0.7646 & 0.1666 & 1.4362 \\
        CPG & 0.8825 & 0.1603$^\ns$ & 2.1836 \\
        DD & 0.7264 & 0.2518 & 2.4382 \\
        DMC & 1.1218 & 0.5568 & 1.8945 \\
        DNL & 1.2014 & 0.1748$^\ns$ & 2.1834 \\
        EEI & 0.8138 & 0.1628$^\ns$ & 2.1700 \\
        EW & 1.0218 & 0.2420 & 1.8502 \\
        FGEN & 0.7343 & 0.3185 & 1.8883 \\
        FLI & 0.9935 & 0.2374 & 1.5466 \\
        FNI & 1.1952 & 0.3580$^\ns$ & 6.0388 \\
        FPH & 0.7587 & 0.0554$^\ns$ & 1.1902 \\
        GERI & 1.0304 & 0.1669$^\ns$ & 2.0919 \\
        GLO & 0.4331 & 0.6612 & 1.8105 \\
        GMA7 & 0.4239 & -0.0013$^\ns$ & 1.3740 \\
        GTCAP & 0.9541 & 0.7777 & 1.8355 \\
        HOUSE & 0.4075 & -0.0132$^\ns$ & 1.7982 \\
        ICT & 0.8903 & 0.7307 & 1.8127 \\
        IMI & 1.1230 & -0.0340$^\ns$ & 2.4140 \\
        JFC & 1.0451 & 0.5168 & 1.7643 \\
        JGS & 1.0386 & 1.1688 & 1.7109 \\
        LC & 0.3592 & -0.2497$^\ns$ & 3.0048 \\
        \hline
    \end{tabular}
    }
    \end{minipage}\hfill
    \begin{minipage}{0.5\linewidth}
    \centering
    \scalebox{0.9}{
    \begin{tabular}{|c|ccc|}
        \hline
        Stock & $\beta_{1i}$ & $\beta_{2i}$ & $\sigma_{i}$ \\
        \hline
        LPZ & 0.8387 & 0.0945$^\ns$ & 2.0117 \\
        LTG & 0.8529 & 0.3950 & 2.0992 \\
        MAXS & 1.2711 & 0.1540$^\ns$ & 1.9789 \\
        MBT & 0.9099 & 0.6560 & 1.5007 \\
        MEG & 1.2877 & 0.7545 & 1.7593 \\
        MER & 0.4251 & 0.6250 & 1.4782 \\
        MPI & 0.9654 & 0.7626 & 1.9291 \\
        MWC & 0.7678 & 0.1878$^\ns$ & 2.2833 \\
        MWIDE & 1.1309 & 0.0369$^\ns$ & 2.3085 \\
        NI & 0.5448 & -0.1908$^\ns$ & 2.4880 \\
        NIKL & 1.3530 & 0.2689$^\ns$ & 3.6684 \\
        PCOR & 0.9715 & 0.1949$^\ns$ & 2.0778 \\
        PGOLD & 0.7043 & 0.2572 & 1.6095 \\
        PLC & 1.1682 & 0.2474 & 2.4584 \\
        PNB & 0.8758 & 0.0977$^\ns$ & 1.5576 \\
        PNX & 0.3580 & 0.2852 & 1.9598 \\
        PX & 0.6711 & -0.0153$^\ns$ & 2.4571 \\
        PXP & 1.2902 & 0.3148$^\ns$ & 5.2036 \\
        RLC & 1.1662 & 0.9296 & 1.9170 \\
        RRHI & 0.6137 & 0.4181 & 1.7350 \\
        SCC & 0.9616 & 0.2731$^\ns$ & 2.8422 \\
        SECB & 0.9898 & 0.4194 & 1.7240 \\
        SLI & 0.3454 & 0.0164$^\ns$ & 2.0057 \\
        SM & 0.7241 & 1.0227 & 1.5732 \\
        SMC & 0.6562 & 0.3536 & 1.7768 \\
        SMPH & 0.8483 & 1.0137 & 1.5017 \\
        SSI & 1.3420 & 0.5626 & 2.7782 \\
        SSP & 0.9091 & 0.1473$^\ns$ & 3.6650 \\
        STI & 0.7093 & 0.1038$^\ns$ & 2.6003 \\
        TECH & 0.9719 & 0.2527$^\ns$ & 3.6977 \\
        TEL & 0.5027 & 0.7467 & 1.8468 \\
        UBP & 0.2440 & -0.0131$^\ns$ & 1.0038 \\
        URC & 0.6913 & 0.7876 & 1.7318 \\
        VITA & 1.1600 & 0.2145$^\ns$ & 3.5918 \\
        VLL & 1.2861 & -0.0188$^\ns$ & 2.0218 \\
        WEB & 1.0343 & 0.2561$^\ns$ & 4.8515 \\
        \hline
    \end{tabular}
    }
    \end{minipage}    
\end{table*}

In summary, the results from both DFM ($n = 1$, $p = 3$, $q = 5$) and DFM ($n = 2$, $p = 2$, $q = 5$) not only align with established theories but also provide a new and alternative understanding to price movement dynamics. In particular, the relationship between the common factors $\boldsymbol{F_t}$ and the PSEi returns provides unique insights into how these models quantify systematic and idiosyncratic movements in the market. Notably, DFM ($n = 2$, $p = 2$, $q = 5$) offers a more nuanced characterization of market dynamics, decomposed as comprising market trend and market volatility, complementing established portfolio risk management theories. Furthermore, the correlation between the factor loadings $\boldsymbol{\beta_i}$ and the CAPM beta highlights the models closely capturing the conventional measure of systematic risk exposure. Overall, the results demonstrate the ability of dynamic factor analysis to provide novel insights into classical market theories.

\section{Nowcasting GDP Application}
The results from the previous section suggest that the common factors $\boldsymbol{F_t}$ are viable real-time market indicators that can be effectively extracted from real-time stock price returns data. These factors hold the potential for extension into various economic and financial applications. This section illustrates the utility of the common factors $\boldsymbol{F_t}$ of the DFM $(n = 2, p = 2, q = 5)$ within an economic context.

One major problem economic leaders face is the substantial delay in the release of key economic indicators. For instance, policymakers rely on the quarterly gross domestic product (GDP) growth rates as the primary indicator of economic performance since it measures the total monetary value of all goods and services produced in a country over a specific period. However, this figure is typically released several weeks after the end of a quarter. During this period, fiscal and monetary policies must be enacted based on incomplete information. To address this issue, many institutions have developed nowcasting models to predict these economic indicators to aid economic decisions. Hence, inspired by previous works \cite{luciani_nowcasting_2018,hayashi_nowcasting_2022,babiiMachineLearningTime2022,ashwinNowcastingEuroArea2021,giannoneNowcastingGDPInflation2008}, this section considers the problem of nowcasting the Philippine GDP growth rates. 

\begin{figure}[h!]
    \centering
    \includegraphics[scale=0.45]{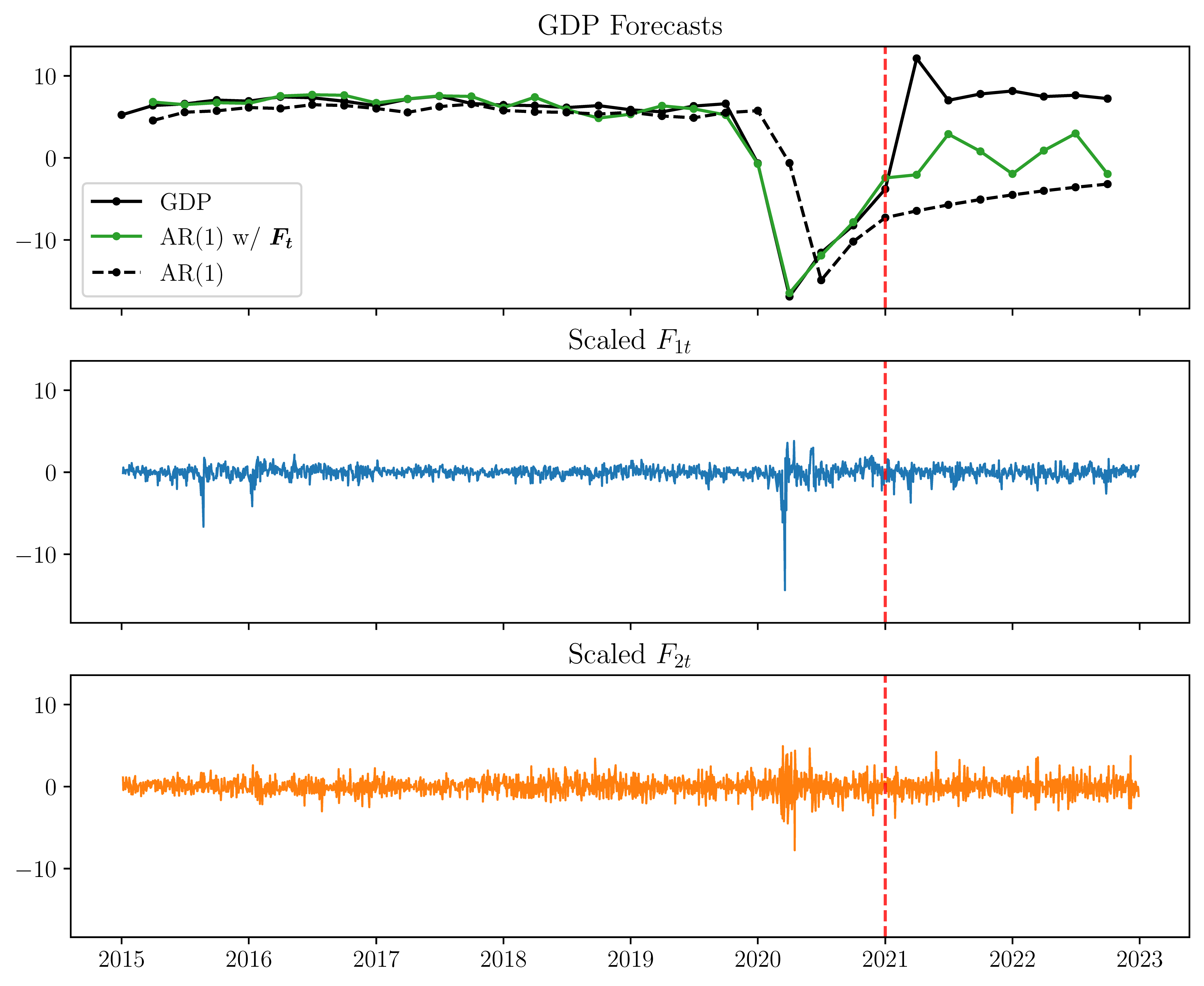}
    \caption{Philippine GDP growth rate nowcasts of AR(1) and AR(1) with $\boldsymbol{F_t}$.}
    \label{fig:resGDP}
\end{figure}

As a benchmark, the AR(1) model is considered. The period from January 1, 2015, to December 31, 2020, is treated as the in-sample period for model training. The period from January 1, 2021, to December 31, 2022, is then treated as the out-of-sample period for model evaluation. Fig.~\ref{fig:resGDP} presents the nowcasts of the model while Table~\ref{tab:rmse} presents a summary of the root mean squared error (RMSE) figures. The model obtained an RMSE of 3.8533 during the in-sample period and 12.3095 during the out-of-sample period. It is worth noting that the out-of-sample RMSE is expected to be larger due to larger fluctuations in GDP growth rates caused by the coronavirus pandemic of 2020. 

\begin{table*}[h!]
    \centering
    \caption{Philippine GDP growth rate nowcasts RMSE values of AR(1) and AR(1) with $\boldsymbol{F_t}$.}
    \label{tab:rmse}
    \setlength{\tabcolsep}{0.6em}
    \renewcommand{\arraystretch}{1.25}
    \scalebox{0.9}{
    \begin{tabular}{|c|cc|}
        \hline
        Model & In-Sample & Out-of-Sample \\
        \hline
        AR(1) & 3.8533 & 12.3095 \\
        AR(1) w/ $\boldsymbol{F_t}$ & 0.6122 & 8.0667 \\
        \hline
    \end{tabular}
    }
\end{table*}

To demonstrate the utility of the DFM, the common factors $\boldsymbol{F_t}$ are first transformed into monthly indicators to adjust for the difference in frequency with the GDP releases. To avoid information leakage, the common factors $\boldsymbol{F_t}$ are the Kalman smoothed common factors during the in-sample period and the Kalman filtered common factors during the out-of-sample period. For each month, the mean and standard deviation of the common factors $F_{1t}$ and $F_{2t}$ are calculated. These monthly indicators are then integrated into the AR(1) model via ordinary least squares regression to explain the GDP growth rate for the quarter. For example, the common factors $\boldsymbol{F_t}$ for January, February, and March are used to explain the GDP growth rate for the first quarter. Fig.~\ref{fig:resGDP} presents the nowcasts of the model while Table~\ref{tab:rmse} presents a summary of the RMSE figures. The model obtained an RMSE of 0.6122 during the in-sample period and 8.0667 during the out-of-sample period. The inclusion of the derived monthly indicators, constructed from the common factors, substantially improved the in-sample RMSE by 84.11\% and the out-of-sample RMSE by 34.47\%, demonstrating the utility of the model. This further supports how the common factors $\boldsymbol{F_t}$ may be used as real-time market indicators.

\section{Conclusion}
In summary, the paper explores the dynamic factor model to analyze price movements in the Philippine Stock Exchange in a manner that combines the predictive capabilities of machine learning models with the interpretability of traditional linear asset pricing models, effectively bridging econometric theory and financial practice. Specifically, the paper focuses on the extracted loadings and common factors to provide alternative perspectives for understanding the dynamics of price movements. Through a validation analysis with the CAPM, the results reveal novel insights into market phenomena. In particular, the one-factor model closely captures systematic or market dynamics similar to the composite index while the two-factor model further decomposes it into market trend and volatility, providing novel perspectives beyond conventional portfolio risk management theories. Additionally, an application on nowcasting GDP growth rates demonstrates the viability of the common factors as real-time market indicators in economic and financial applications by providing substantial performance improvements. These results assert the value of dynamic factor analysis in providing a deeper understanding of price movements in the market.  

Future studies may build on the current findings by relaxing the assumptions of the DFM in Gaussian stock price returns and cross-sectional independence in the errors $Z_{it}$. Specifically, it may be useful to explore deviations from normality in the form of heavy tails as observed in empirical results \cite{Peiró01121994,li2023explanationdistributioncharacteristicsstock}. To this end, one might incorporate a generalized autoregressive conditional heteroskedasticity (GARCH) process for the common factors $\boldsymbol{F_t}$ to allow for time-varying conditional variances, leading to more accurate modeling of stock return distributions. While this would typically introduce more layers of complexity to the model-fitting methodology, they may be alternatively accommodated through approximate inferential methods utilizing advances in variational inference algorithms \cite{dayta2024you}. Additionally, replacing the Kalman filter for model fitting with variational inference may allow for higher orders of flexibility in terms of the factor distribution and other additional effects, such as GARCH terms and market or industry-level dynamics. These model improvements would better capture the intricate dynamics of stock price movements while maintaining their predictive performance and interpretability.

\section*{Declarations}
\subsection*{Availability of data and materials}
The datasets generated and analyzed during the current study are available in the repository \url{https://github.com/briangodwinlim/DynamicFactorAnalysis}.

\subsection*{Competing Interests}
The authors have no competing interests to declare.

\subsection*{Funding}
The Japan Society for the Promotion of Science supported this study through Grants-in-Aid for Scientific Research Program (KAKENHI 18K19821).

\subsection*{Author's Contribution}
BGL, DD, and BRT contributed to the overall conceptualization, code development, data gathering, statistical analyses, and manuscript writing. RRT, LPDG, and KI contributed to the overall conceptualization, manuscript writing, and manuscript finalization. All authors have read and approved the final version of the manuscript.

\subsection*{Acknowledgements}
Not applicable.

\subsection*{Author's Information}
BGL, DD, and BRT are doctoral students at the Division of Information Science, Nara Institute of Science and Technology. RRT is an assistant professor at the Division of Information Science, Nara Institute of Science and Technology, a research fellow at the Graduate School of Informatics, Kyoto University, and an affiliate faculty at the John Gokongwei School of Management, Ateneo de Manila University. LPDG is a lecturer at the School of Mathematical and Physical Sciences, University of Technology Sydney, and an affiliate faculty at the School of Science and Engineering, Ateneo de Manila University. KI is a professor at the Division of Information Science, Nara Institute of Science and Technology.

\section*{Abbreviations}
\begin{tabular}{ll}
    CAPM & Capital Asset Pricing Model \\
    APT & Arbitrage Pricing Theory \\
    PCA & Principal Component Analysis \\
    DFM & Dynamic Factor Model \\
    VAR & Vector Autoregressive Process \\
    AR & Autoregressive Process \\
    PSEi & Philippine Stock Exchange Index \\
    BIC & Bayesian Information Criterion \\
    GDP & Gross Domestic Product \\
    RMSE & Root Mean Squared Error \\
    GARCH & Generalized Autoregressive Conditional Heteroskedasticity \\
\end{tabular}

\bibliography{sn-bibliography}

\end{document}